\documentclass[conference]{IEEEtran}
\IEEEoverridecommandlockouts
\usepackage[hidelinks]{hyperref}
\usepackage{algorithm,algpseudocode}
\usepackage{tabularx}
\usepackage{url}
\usepackage{amsmath,amssymb,amsfonts}
\usepackage{algpseudocode}
\usepackage{graphicx}
\usepackage{textcomp}
\usepackage{comment}
\usepackage{booktabs}
\usepackage{caption,subcaption}
\usepackage{geometry}
\usepackage{pdfpages}
\usepackage{soul}
\usepackage{booktabs}
\newcolumntype{W}{>{\hsize=20\hsize}X}
\newcolumntype{Y}{>{\centering\arraybackslash}X}
\newcolumntype{Z}{>{\hsize=.2\hsize\raggedleft\arraybackslash}X}
\def\BibTeX{{\rm B\kern-.05em{\sc i\kern-.025em b}\kern-.08em
    T\kern-.1667em\lower.7ex\hbox{E}\kern-.125emX}}
\usepackage[
backend=biber, style=numeric, sorting=none, maxnames = 50
]{biblatex}

\newcommand{\reviewerone}[1]{\textcolor{blue}{#1}}

\newcommand{\example}[1]{{\tt #1}}
\newcommand{\centered}[1]{\begin{tabular}{l} #1 \end{tabular}}

\begin{document}

\title{Methods for Matching English Language Addresses\\
%{\footnotesize \textsuperscript{*}Note: Sub-titles are not captured in Xplore and should not be used}
%\thanks{Identify applicable funding agency here. If none, delete this.}
}

\author{\IEEEauthorblockN{Keshav Ramani}
\IEEEauthorblockA{
\textit{J.P. Morgan AI Research}\\
London, UK \\
keshav.ramani@jpmchase.com
}
\and
\IEEEauthorblockN{Daniel Borrajo$^1$\thanks{$^1$On leave from Universidad Carlos III de Madrid}}
\IEEEauthorblockA{
\textit{J.P. Morgan AI Research}\\
Madrid, Spain \\
daniel.borrajo@jpmchase.com
}
}

\maketitle

\begin{abstract}
Addresses occupy a niche location within the landscape of textual data, due to the positional importance carried by every word, and the geographical scope it refers to. The task of matching addresses happens everyday and is present in various fields like mail redirection, entity resolution, etc. Our work defines, and formalizes a framework to generate matching and mismatching pairs of addresses in the English language, and use it to evaluate various methods to automatically perform address matching. These methods vary widely from distance based approaches to deep learning models. By studying the Precision, Recall and Accuracy metrics of these approaches, we obtain an understanding of the best suited method for this setting of the address matching task.
\end{abstract}

\begin{IEEEkeywords}
address-matching, deep learning, natural language processing, embedding
\end{IEEEkeywords}

\section{Introduction}
One of the most interesting types of naturally occurring text are addresses. Fundamentally, these text types denote the location of an entity by specifying the hierarchy of which geographic entities they are a part of. These entities can be countries, states, cities, districts, roads, streets, buildings and more. Universally there is no strict definition of what fields must be included in a string for it to qualify as an address, but it is typically a combination of some or all of the aforementioned attributes. Different countries, and states have their own conventions of how addresses must be structured. However, it is safe to say that they are generally a sequence of words and numbers arranged in an increasing order of geographical scope, starting from apartment, floor and/or building information all the way to which country they are present in.

In this paper we deal with the task of address matching, whose compact formulation is: given the representations of two addresses, return true if the two representations refer to the same address. This task can take a very simplified version when addresses are represented as sets of components (apartment, floor, building, street name, city, post code, state, country, \ldots). In these cases, some simple rules can be defined to check whether they refer to the same country, state, city and so on.
However, in many real world cases, the addresses are just represented as a single string. Thus, the task of matching addresses involves looking at two strings and evaluating whether they refer to the same geographic location (under a given a level of granularity).

Addresses appear in a wide variety of documents. Some potential applications of address matching are in the field of mail routing, clustering of addresses and matching a given list of addresses of interest to a larger database. In the case of postal address routing, address matching could be required to work at the granularity of the person/property involved. When it comes to clustering of addresses for purposes like grouping of households, communities, etc\reviewerone{,} the granularity could range anywhere between building level to district levels.

Consider the following pairs of addresses:

\begin{table}[ht]
\centering
\begin{tabular}[t]{lcc}
\toprule
Pair \#&Address 1&Address 2\\
\midrule
1&123 ABC Court&123 ABC Ct\\
2&123 ABC Court&23 ABC Court\\
3&123 ABC Court&123z ABC Court\\
4&123 ABC Court&ABC Court\\
\bottomrule\\
\end{tabular}
\caption{Some examples of the difficulty of address matching.}
\label{tab:examples}
\end{table}%

Let us consider two addresses to be a match if they are referring to the same building. By this assumption, pair 1 has a very high chance of being a match, given that in countries like the USA, "Court" is often abbreviated to "Ct". Pair 2 clearly isn't a match, given that they refer to different locations within ABC Court. Pair 3 is interesting as one might assume that the 'z' at the end of 123 is an erroneous character, or that '123z ABC Court' is physically present within '123 ABC Court'. Despite Pairs 2 and 3 having an edit distance of 1, we see that the  chances of them being treated as similar addresses can greatly differ. Pair 4 is once again interesting, as Address 1 could be present at Address 2.

We can see that the level of granularity and a specific application might influence the labelling process. However, irrespective of the nuanced examples and domain relevant subtleties, one can clearly start seeing why this problem is difficult: 

\begin{itemize}
    \item Addresses are a unique subset of naturally occurring text. Conventional methods of string matching will not generally be effective here.
    \item There is a unique subconscious method humans employ to match addresses that has not been imitated thoroughly by a computer yet. While there is some literature that has studied this question from a computational perspective (further described in Section~\ref{sec:related-work}), it is nowhere as standardized or thoroughly investigated as other sub-classes of natural language tasks (literature, dialogue, news).
\end{itemize}

This paper investigates the question of whether the process of address matching can be automated, especially by exploiting recent advancements in the field of Natural Language Processing. Our first contribution is an address matching task generator that allows us to automatically generate tasks of varying difficulty (Section~\ref{sec:datset}). The second contribution is the development of a suite of techniques to solve the task (Section~\ref{sec:algos}). Among them, we propose a novel algorithm based on current deep learning techniques. The third contribution is an experimental comparison and analysis of results among the different techniques (Section~\ref{sec:experiments}). We finalize the paper with some conclusions  (Section~\ref{sec:conclusions}).

\section{Prior Work}
\label{sec:related-work}

Prior research works have approached this problem using multiple techniques with varying degrees of sophistication. However, the progress of building intelligent systems to solve this task differs across languages. For example, address matching in Mandarin Chinese has seen a considerable degree of improvement when compared to English~\cite{lin20}~\cite{comber2019machine}. 

As far as matching addresses in the English language go, Santos \textit{et al.} explored the idea of matching toponyms using various methods~\cite{santos2018toponym}. Comprehensively, they analyze the effectiveness of using various string distance metrics, straightforward machine learning approaches like SVMs \cite{boser1992training} \cite{cristianini2000introduction}, and even build a deep learning system for this task. While this work does not exactly deal with addresses (toponyms are typically smaller strings that are equivalent ways of addressing cities/islands), it sufficiently motivates the study of the address matching problem from a computational viewpoint.

The work of Comber \textit{et al.}~\cite{comber2019machine} however, focuses more directly with the problem of address matching. They primarily use Conditional Random Fields (CRF) to segment the addresses. They argue that unlike previously used techniques like HMMs which assume independence between new labels and previously predicted labels, CRF are conditional by nature and thus make use of previous information. Once they have segmented the addresses, they proceed to match them using blocking techniques. Another approach they pursue is to use word2vec embeddings~\cite{mikolov2013efficient} to augment the representations generated by the CRF. While this work is highly innovative, it is not possible for the blocking module to provide feedback to the CRF so that it may enrich the representations specifically for the task of matching addresses.

The work of Lin \textit{et al.,} ~\cite{lin20} made two significant contributions (from which this work is also inspired): an address matching corpus in Mandarin, and a deep address matching algorithm. From a given dataset of unique addresses, the authors were able to apply string transformations to generate candidate matching addresses. Mismatches were also generated by simply choosing another address. Thus, for each address a match and a mismatch address were generated. Once this corpus was created, the authors were able to train an ESIM model~\cite{Chen_2017} to predict whether a given pair of addresses were a match or a mismatch. A few ways in which our work differs from theirs is that their model mainly relies on word2vec to provide embeddings, while the deep learning model we use employs Glove embeddings~\cite{pennington2014glove}. Secondly, their address dataset is limited to one city in China, and therefore does not encounter various roads/buildings of the same name that span across multiple cities. Our dataset is currently based on the structure of addresses found in the USA. Thirdly, our work has a few improvements to address generation, in that the mismatches we generate can also be more nuanced - by modifying the property/road numbers (discussed further in the next section). Finally, we also study the effect of adding character embeddings to the ESIM model, within this problem setting.

\section{Dataset Generation}
\label{sec:datset}

Some of the techniques described in the next section to solve the address matching task are based on machine learning. Thus, we would first need a dataset to train learning algorithms, as well as to evaluate all address matching algorithms. Though previous works have used datasets for similar tasks, they have either been in a different language or they are not available~\cite{chen2021deep}~\cite{lin20}. Thus, we have generated a dataset for this task, which is inspired by the approach that Lin {\it et al.} followed in their work~\cite{lin20}.

A prerequisite to generating the data is a clear understanding of what exactly constitutes the task of matching two addresses. At the finest granularity, we may define that two addresses match if they refer to the same exact residential/official unit. This would mean that while two locations share the same geographic physical location (say latitude and longitude co-ordinates), they could be two different apartments in the same building and thus could be considered mismatching addresses. Other applications might define matching addresses to be a pair of addresses that refer to the same physical structure, say a building. This is sensible in applications where there is a common mailbox, or institutions in the same physical location that need to be grouped together. Within this paper, we define matching addresses to be a pair of addresses that refer to the same building. The main reason to choose this definition is that we are primarily interested on financial services applications, and many of those applications share this definition. Nevertheless, all algorithms presented in the next section are agnostic to this definition, and we could have used a definition at a lower or greater level of granularity.

The process of generating the matching addresses dataset is composed of four steps: base address generation, prefix generation, matching address generation, and mismatching address generation. These three steps are described next.

\subsection{Base Address Generation} 

In order to generate this dataset, we first generated 10,000 base addresses. A base address is represented as a string, and follows the structure:

\begin{center}
    $<$Building, Street Name, City, State$>$
\end{center}

For each base address, the building number and the street name were generated using the Faker synthetic data generator\footnote{https://github.com/joke2k/faker}. The cities were randomly drawn from the top 1000 cities by population in the USA. This was extracted from a public dataset provided by plotly\footnote{https://github.com/plotly/datasets/blob/master/us-cities-top-1k.csv}. Care was taken to preserve the mapping between these cities and the relevant states they were present in, as we did not want spurious mappings to be taught to the model. 

\subsection{Prefix Generation} 

Given that our work focuses on matching addresses at the building level, a crucial piece of information that is missing from the base addresses is the exact location of the entity being addressed within the building. In order to make the dataset generic for other definitions of address matching, this information can be provided by various factors like the floor/apartment/suite in which the entity is present. In some cases, even names of people are used directly. For a given address, the type of prefix is chosen randomly between Floor, Apartment, and Name. For floor and apartment prefixes, the word is drawn from the relevant lists shown on Table~\ref{tab:prefix}\reviewerone{,} and a random number is assigned along with it. In the case of names, we first draw a prefix from the list mentioned in the same table\reviewerone{,} and a random name is assigned along with it. The name is generated using the Faker library.

\begin{table}[ht]
\centering
\begin{tabular}[t]{cr}
\toprule
{\bf Group} & \multicolumn{1}{c}{\bf Prefix List}\\
\midrule
Apartment& APT, APARTMENT, SUITE, STE, UNIT\\
Floor&FLOOR, LEVEL\\
Name&ATTN, C/O\\
\bottomrule
\end{tabular}
\caption{Word equivalences used for generating addresses.}
\label{tab:prefix}
\end{table}%

This process of prefix generation will be used downstream to generate matching and mismatching addresses. 

\subsection{Matching address generation}

The target dataset that we aspire to generate is comprised of matching and mismatching address. The transformations involved in generating said matches or mismatches must be capable of capturing nuances observed in the real world. We shall now describe the string transformations that are used to generate a matching address from a given address.

\subsubsection{Word Substitution}
\label{AA}

Certain words in addresses can be described in different ways. A model that must capture similar addresses needs to be trained in recognizing such pairs. Table~\ref{tab:word_substitutions} captures the substitutions being considered in our formulation. Thus, we define a substitution operation as the process of replacing one word from a given group by another word from the same group that is randomly selected.

\begin{table}[htb]
\centering
\begin{tabular}[t]{cr}
\toprule
{\bf Group} & \multicolumn{1}{c}{\bf Words}\\
\midrule
1& APT, APARTMENT\\
2&SUITE, STE\\
3&ROAD, RD\\
4&STREET, ST\\
5&AVE, AVENUE\\
\bottomrule
\end{tabular}
\caption{Word substitutions used to modify addresses.}
\label{tab:word_substitutions}
\end{table}%

\subsubsection{Word Deletion}
\label{BB}
One of the common errors involved in handling addresses consists of dropping words. Thus, we define the process of word deletion as a word from an address being randomly dropped. We do not drop words that could constitute the name/street name/building number in an address. Dropping these fields fundamentally would change who/which building the address references.

\subsubsection{Character Addition}
\label{CC}
Noisy characters sometimes tend to seep into addresses, and it would be expected from the model to be robust to minor spelling errors. Thus, we define a character addition operation that includes a random ASCII character at a random location within the street name. We do not opt to add a random character in other locations of the address, as it might refer to a different address. \example{123 ABC Ct} is different from \example{123z ABC Ct}.

\subsubsection{Character Deletion}
\label{DD}
Similar to how we would expect the model to be robust to the effects of character addition, we would expect it to be robust to character removal as well. For this operation we once again only consider the street names. Deleting characters in other fields might lead to potentially different addresses being matched together. \example{123 ABC Ct} is different from \example{12 ABC Ct}.

\subsubsection{Permutation}
\label{EE}
In digital applications as well as fields like post delivery, it is not uncommon to find the first and second lines of addresses to be swapped. This can be attributed to differing conventions in various applications/countries.

In order to incorporate this effect, we implement the permutation operation where we switch the order of occurrence of certain fields within an address. Specifically, if a given address has a floor prefix (like \example{FLOOR} or \example{LEVEL}) followed by a number, the permutation operation swaps its occurrence with the building number and street name. \example{FLOOR\ 3\ 123\ ABC\ Ct} gets transformed to \example{123\ ABC\ Ct\ FLOOR\ 3}.

In addition to floor information, this operation also has the same effect on apartment information, or any other prefix. Thus if a given address has an apartment prefix (\example{APARTMENT}, \example{SUITE}, \example{UNIT}) followed by a number, the operation swaps it with the building number and state name. \example{APARTMENT\ 15\ 123\ ABC\ Ct} gets transformed to \example{123\ ABC\ Ct\ APARTMENT\ 15}

\subsection{Mismatch generating transformations}
Having described the matching address generation transformations, we will now proceed to describe the various transformations used to generate mismatching addresses. It is important to note that these transformations are capable of capturing subtle and crucial modifications that can elicit a mismatching address, as well as obvious mismatches.

\subsubsection{Building Redirection} 
\label{FF}
One of the nuances of mismatches that were earlier alluded to in the Introduction is that the model must be able to understand the significance of a change in the numerals in the address. Though the string edit distance between \example{123 ABC Ct} and \example{23 ABC Ct} is just 1, the model must understand that they refer to two different addresses. Further, when assessed through the impact of a pair of addresses being considered a match or a mismatch, this type of a difference is fundamentally different from a spelling error. Thus, we define building redirection as an operation that changes the building number in an address by adding to or subtracting values from it. 

\subsubsection{Street Redirection}
\label{GG}
A coarser form of redirection consists of altering the street name present in an address. For this operation, we use the Faker library to replace an existing street name with a new street name. A suitable example for this is \example{123 ABC Ct} could get converted to \example{123 XYZ St}, where \example{XYZ St} is a street name arbitrarily generated by the Faker library.

\subsubsection{City Redirection}
\label{HH}
The coarsest form of redirection we consider is to alter the city name in an address. A change in the city will result in a potential change to the state mentioned in the address as well. For this operation, we replace the city and state by randomly sampling a city state pair from the list of cities previously mentioned. This follows from the fact that it is common to have two matching street level addresses in two completely different cities. A suitable example for this could be \example{123 ABC Ct Edison NJ} potentially getting converted into \example{123 ABC Ct Atlanta GA}

\subsection*{\textbf{Generation of Matching/Mismatching Address Pairs}}

Given the previous operations, we generate matches and mismatches, as positive and negative examples.

\textbf{Match Generation: }From a given base address, we first obtain two different addresses by applying the prefix generation function twice. At this point, the addresses are still matching as they refer to the same building, but have differing prefixes. We further increase the linguistic differences between these two addresses by applying one of the operations from \ref{AA}, \ref{BB}, \ref{CC}, \ref{DD}, and \ref{EE}. This choice is made randomly for each address in the pair. All these operations do not alter the building being addressed.

Figure~\ref{fig:matching_trans} shows a couple of match generation scenarios. It is important to note that the operations are assumed to be chosen randomly. $A1$ and $A2$ refers to the two addresses generated by adding prefixes to the corresponding base address.

\begin{figure}[hbt]
\hrule \relax
\begin{center}
    \textbf{Base Address:}\hspace{12mm} \example{123\ ABC\ CT\ LIMA\ OH}
    \newline
    
    \textbf{Adding new prefixes:}\hspace{41mm}
    \newline
    A1:\hspace{16mm} \example{APT\ 3\ 123\ ABC\ CT\ LIMA\ OH}
    \newline
    A2:\hspace{14mm} \example{STE\ 17\ 123\ ABC\ CT\ LIMA\ OH}
    \newline
    \newline
    \textbf{Applying matching transformations:}\hspace{18mm}
    \newline
    A1 + \ref{EE}:\hspace{3mm} \example{123\ ABC\ CT\ APT\ 3\ LIMA\ OH}
    \newline
    A2 + \ref{BB}:\hspace{7mm} \example{STE\ 123\ ABC\ CT\ LIMA\ OH}
\end{center}

\begin{center}
    \textbf{Base Address:}\hspace{12mm} \example{123\ ABC\ CT\ LIMA\ OH}
    \newline
    
    \textbf{Adding new prefixes:}\hspace{41mm}
    \newline
    A1:\hspace{10mm} \example{ATTN\ JOE\ 123\ ABC\ CT\ LIMA\ OH}
    \newline
    A2:\hspace{12mm} \example{UNIT\ 12\ 123\ ABC\ CT\ LIMA\ OH}
    \newline
    \newline
    \textbf{Applying matching transformations:}\hspace{18mm}
    \newline
    A1 + \ref{BB}:\hspace{7mm} \example{JOE\ 123\ ABC\ CT\ LIMA\ OH}
    \newline
    A2 + \ref{AA}:\hspace{1mm} \example{UNI\ 12\ 123\ ABC\ CT\ LIMA\ OH}
    \newline
\end{center}

\hrule \relax
\caption{Examples of matching address generation with randomly chosen transformations.}
\label{fig:matching_trans}
\end{figure}

\textbf{Mismatch Generation: } We start off similar to match generation by generating two different addresses from a given base address by using different prefixes. On one of these addresses, we now apply an operation randomly chosen out of \ref{FF}, \ref{GG}, \ref{HH} or randomly sampling another base address. The geographic scope of these operations, as well as sampling another base address, is between building number level to state level impact. The output of these operations, therefore, will fundamentally produce a mismatch to the other address in the pair. 

While this address generator was tailor made for the task of matching addresses at a building level, one may also re-purpose this to extend it to tasks outlined in Section 1. For example, if we would like to match addresses at the level of names, then we will need the addresses to be more textually similar than our current use case. This would mean that for matching addresses, we would still be allowed to perform word substitutions for name prefixes. However, character additions or deletions become a mismatch generation operation at this granularity. For a different use case, say district level clustering, various replacements and substitutions can be incorporated into the name and street name fields. However, we will once again need to corrupt the district names to generate mismatches. The operations defined here are general and can be easily be adapted for these use cases.

Figure~\ref{fig:mismatching_trans} shows two examples of how mismatching addresses are being generated. The transformations used have been chosen randomly.

\begin{figure}[hbt]
\hrule \relax
\begin{minipage}{0.5\textwidth}
    \textbf{Base Address:} \example{123\ ABC\ CT\ LIMA\ OH}
    \newline
    
    \textbf{Adding new prefixes:}
    \newline
    A1: \example{APT\ 3\ 123\ ABC\ CT\ LIMA\ OH}
    \newline
    A2: \example{STE\ 17\ 123\ ABC\ CT\ LIMA\ OH}
    \newline
    \newline
    \textbf{Applying mismatching transformations:}
    \newline
    A1 + \ref{FF}: \example{APT\ 3\ 124\ ABC\ CT\ LIMA\ OH}
    \newline
    A2 + \ref{HH}: \example{STE\ 17\ 123\ ABC\ CT\ RENO\ NV}
    \newline
\end{minipage}

\begin{minipage}{0.6\textwidth}
    \textbf{Base Address:} \example{123\ ABC\ CT\ LIMA\ OH}
    \newline
    
    \textbf{Adding new prefixes:}
    \newline
    A1: \example{ATTN\ JOE\ 123\ ABC\ CT\ LIMA\ OH}
    \newline
    A2: \example{UNIT\ 12\ 123\ ABC\ CT\ LIMA\ OH}
    \newline
    \newline
    \textbf{Applying mismatching transformations:}
    \newline
    A1 + \ref{FF}: \example{JOE\ 124\ ABC\ CT\ LIMA\ OH}
    \newline
    A2 + \ref{GG}: \example{\color{blue}UNIT\color{black}\ 12\ 123\ XYZ\ CT\ LIMA\ OH}
    \newline
\end{minipage}

\hrule \relax
\caption{Examples of mismatching address generation with randomly chosen transformations.}
\label{fig:mismatching_trans}
\end{figure}

\subsection{Dataset Generation: }The previously mentioned 10,000 base addresses are passed to the Match generators and Mismatch generators. Each of these result in 10,000 pairs of matching and mismatching addresses. These records are then shuffled and 80\% of the data is used to construct the training set, 10\% for the validation and the remaining 10\% for the test set.

\section{Address Matching Algorithms}
\label{sec:algos}

Having constructed a dataset of matching addresses tasks, we now proceed to describe the various algorithms and models we used for this task of address matching.
Given that addresses differ from generic natural language sentences in a small but significant way, the candidate models to compare against are vast. In our work, we have considered various string distance based approaches, and a modified version of the state of the art model for addresses in Mandarin proposed by Lin \textit{et al.}~\cite{lin20}. The following subsections will be diving deeper into each of these approaches.

\subsection{Baseline Algorithms}
One of the most common approaches taken to solve the problem of address matching is to treat addresses as generic texts, create structured representations, and apply some kind of similarity metric to the structured representations. So, we can create a suite of baseline algorithms by varying the structured representation and the similarity functions.

Upon receiving an input address in the form of a string, the first step consists of creating a dictionary with a single key \example{Address} that contains a list with a single element with the original string as value. The values of the dictionary are defined as lists given that, as we shall see below, some representation changes generate a list of values.

Structured representations are computed by functions $R: D\rightarrow D$. They take as input an address in the form of a dictionary and return as output a new dictionary representing the address. The keys of the dictionary are: \example{Address}, \example{Person}, \example{Unit}, \example{Floor}, \example{House}, \example{Area-District}, \example{POBox}, \example{Street}, \example{StreetNumber}, \example{StreetName}, \example{PostCode}, \example{City}, \example{County-State}, and \example{Country}. The \example{Person} key relates to the fact that some addresses contain information on some person living on that address, as in \example{Attn. John Smith, 123 Main St.}. 

We have implemented variations of standard representations for addresses as the following:

\begin{itemize}
    \item {\bf normalized}: it replaces the \example{Address} value (list of a single string) of the address (dictionary) with a normalized version of the address. It operates over the string by first performing some string clean up operations. For instance, it removes non-ASCII characters, or provides homogeneous use of delimiters (e.g. comma, period, \ldots). Second, it deals with the standard use of abbreviations and equivalences on addresses. Typical examples include word groups in Table~\ref{tab:word_substitutions}. We use a list of 150 terms and their different utterances to normalize the addresses. These terms are divided depending on which field they refer to: \example{Unit} (e.g. apartment, unit, suite, \ldots), ); \example{Floor} (e.g. floor, level, basement, \ldots), \ldots. As an example, an input address such as \example{--- 3rd floor Howard Bldg, 123 W. Main St. --} would be normalized into: \example{3 $\langle$Floor$\rangle$ Howard $\langle$Building$\rangle$, 123 $\langle$West$\rangle$ Main $\langle$Street$\rangle$}.
    
    \item {\bf tokens}: it replaces the \example{Address} value (a list of a single string $s$) with a list of tokens in that string $s$. So, the above example would generate the new value of the \example{Address} as a list composed of: \example{[3, $\langle$Floor$\rangle$, Howard, $\langle$Building$\rangle$, 123, $\langle$West$\rangle$, Main, $\langle$Street$\rangle$]}.
    
    \item {\bf n-grams}: it replaces the \example{Address} value with a list of n-grams corresponding to the strings in the list. We have used 3-grams. So, in case we apply n-grams to the normalized address representation (\example{3 $\langle$Floor$\rangle$ Howard $\langle$Building$\rangle$, 123 $\langle$West$\rangle$ Main $\langle$Street$\rangle$}), we would obtain: \example{['3 $\langle$', '  $\langle$F', '$\langle$Fl', 'Flo', 'loo', 'oor', 'or$\rangle$', \ldots, 'et$\rangle$']}.
    
    \item {\bf tf-idf}: it replaces the \example{Address} value (list of strings) with a list of TFIDF values of those strings~\cite{tfidf}. The strings can be n-grams or tokens computed by the previous representation change functions.
    
    \item {\bf segmentation}: it populates the address dictionary with the above mentioned keys. Their values are captured from the \example{Address} string by using some regular expressions that define the corresponding values. In the case of cities, states and countries, we use lists of those and then automatically create a single regular expression that captures them all. As an example, the previously normalized address, \example{3 $\langle$Floor$\rangle$ Howard $\langle$Building$\rangle$, 123 $\langle$West$\rangle$ Main $\langle$Street$\rangle$}, would generate the dictionary: \example{Floor = '3', Building = 'Howard $\langle$Building$\rangle$', StreetNumber = '123', StreetName = '$\langle$West$\rangle$ Main $\langle$Street$\rangle$'}.
\end{itemize}

Given that these functions take as input and generate as output the same structure, they can be chained. As an example, we can first normalize and then generate n-grams. There are some constraints to these chains. For instance, segmentation is applied first. Furthermore, once tf-idf has been used, we cannot use the other representation changes. And usually, it does not make much sense to use both n-grams and tokens, so we only allow one of them to apply. Finally, normalization usually only makes sense to be applied at start.

With respect to the similarity functions, we have implemented (or used their implementation on scikit-learn~\footnote{https://scikit-learn.org/stable/}) the following functions. They all take as input two dictionaries that represent addresses and return True if they match.

\begin{itemize}
    \item {\bf simple}: it returns true if the values (lists) of all keys match. To check whether two lists match, first they are both converted into sets. They match if the sets are equal. If one of the lists is empty, we assume both lists match.
    
    \item {\bf jacquard}: it returns true if the values (lists) of all keys match. To check whether two values (lists) match, first they are both converted into sets $s1$ and $s2$. They match if Equation\ref{eq1} holds.
    
    \begin{equation}
    \label{eq1}
    1 - \frac{s1\cap s2}{s1\cup s2} < JT * \min(|s1|, |s2|)
    \end{equation}
    
    where $JT$ is the Jacquard threshold to accept a match. Experimentally, we have found that a value of 0.05 produces the best results.
    
    \item {\bf levenshtein}: it returns true if the values (lists) of all keys match. To check whether two values (lists) of a key $k$ match, it computes the Levenshtein distance for each pair of strings in those two lists. Then, it computes the average of those distances for key $k$, $\mu_{k}$. It then checks if Equation~\ref{eq2} holds for $\mu_{k}$.
    
    \begin{equation}
    \label{eq2}
        \mu_{k} \geq LTV_{\mbox{min}} * ML
    \end{equation}
    
    where $LTV_{\mbox{min}}$ is a parameter that defines the maximum allowed averaged distance percentage for values of keys, and $ML$ is the minimum length of all strings in the two values. If the formula holds, then the values for key $k$ are considered to be far from each other and the addresses are considered not to match. We have experimentally obtained the value 0.2 to have the best performance.
    
    If all values of all keys match, we still make a final check according to Equation~\ref{eq3}.

   \begin{equation}
    \label{eq3}
        \mu \geq LTA_{\mbox{max}} * MA
    \end{equation}
    
    where $\mu$ is the average distance across all keys, $LTA_{\mbox{max}}$ is a parameter that defines the maximum allowed averaged distance percentage for all keys, and $MA$ is the minimum length of the original addresses. If the formula holds, then the addresses are considered not to match. We have experimentally obtained the value 0.2 to have the best performance.
    
    \item {\bf Jaro-Winkler}: instead of using the Leveshtein distance, it uses the Jaro-Winkler distance. The approach is the same as before. The values for the corresponding parameters are\reviewerone{:} $LTV_{\mbox{min}}$=0.5 and $LTA_{\mbox{max}}$=0.002.
    
    \item {\bf cosine}: in the case of representations based on tf-idf, the standard distance  is the cosine function that computes the cosine of the vectors representing the addresses. Given that this function returns values between 0 and 1, this approach requires setting a parameter that is the threshold that establishes which cosine similarity value is the upper bound for defining similarity. Empirically, we found that a value of 0.75 provides good results.
    
\end{itemize}

Given these two functions (representation and similarity), we can define many different valid configurations. In order to characterize the different alternatives, we define them based on several features: normalization, segmentation, tokens, n-grams, tf-idf and distance. All are boolean except for the distance metric that will take as value all the similarity/distance functions defined above. Thus, a simple algorithm could be defined as: \{'normalization': True, 'segmentation': False, 'tokens': False, 'n-grams': False, 'tf-idf': False, 'distance': 'simple'\}. The only representation change would be normalizing the input, and the distance function would be the simple one. Table~\ref{tab:algos} presents the variations we have used in our experiments. An X means the corresponding feature is True, while no value means it is False.
                    
\begin{table*}[hbt]
    \centering
    \begin{tabular}{lcccccl} \toprule
    \multicolumn{1}{c}{algorithm} & normalization & segmentation & tokens & n-grams & tf-idf & \multicolumn{1}{c}{distance}\\ \midrule
plain & & & & &  & simple\\
normalized-plain & X &  &  &  &  & simple\\
tokens-jacquard & X &  & X &  &  & jacquard\\
n-grams-jacquard & X &  &  & X &  & jacquard\\
levenshtein & X &  &  &  &  & levenshtein\\
jaro-winkler & X &  &  &  &  & jaro-winkler\\
tfidf & X &  &  & X & X & cosine\\
segment & X & X &  &  &  & simple\\
segment-levenshtein & X & X &  &  &  & levenshtein\\
segment-jaro-winkler & X & X &  &  &  & jaro-winkler\\
segment-tokens-jacquard & X & X & X &  &  & jacquard\\
segment-n-grams-jacquard & X & X &  & X &  & jacquard\\
segment-tfidf & X & X &  & X & X & cosine\\ 
\bottomrule
\end{tabular}
    \caption{Definition of algorithms used in the experiments. An X means the corresponding feature is True, while no value means it is False.}
    \label{tab:algos}
\end{table*}

\subsection{ESIM Based Model}
The ESIM model, first proposed by Chen \textit{et al.}~\cite{Chen_2017} is a sequential inference model that works with word embeddings, Bi-LSTMs and an attention mechanism. As shown in Figure~\ref{fig:architecure}, the input provided to this model is a pair of strings, and the output it learns is typically a boolean variable.

While the ESIM model typically works with only word embeddings, we studied the effect of adding character embeddings as well. This idea has been previously explored in settings like context response matching in dialgoue corpora \cite{dong2018enhance}. The motivation behind extending this framework to address matching is that in cases where character noise is present, character embeddings can be used to address any errors in tokenization or word level matching. As shown in Figure~\ref{fig:architecure}, for each address inputs, the words and characters are used to generate the embeddings $\overline{w}$ and $\overline{c}$. These two embeddings are concatenated as a combined address representation which is used downstream for other operations. These embeddings are passed to an attention layer to learn the relative importance between the characters and words of both addresses.

\begin{figure}[hbt]
    \centering
    \makebox[\columnwidth][c]{\includegraphics[trim={12.5cm, 6cm, 12.5cm, 6cm}, clip, width=\columnwidth, page=2]{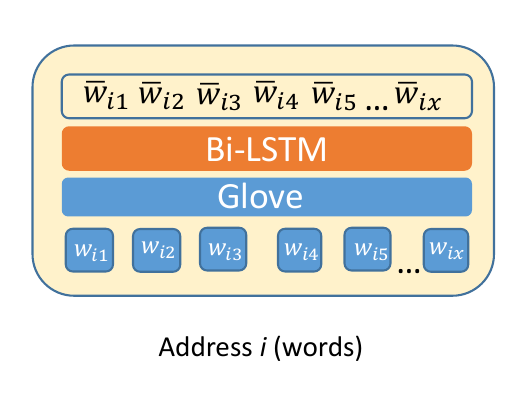}}
    
    \caption{\textbf{Word vector computation.} The words in an address are passed through a glove embedding layer and the outputs from thereon are passed to a Bi-LSTM layer. As Chen \textit{et al.} noted, Glove \cite{pennington2014glove} can be a good choice of word embeddings for the ESIM model.}
    \label{fig:word_vectors}
\end{figure}

\begin{figure}[hbt]
    \centering
    \makebox[\columnwidth][c]{\includegraphics[trim={12.5cm, 6cm, 12.5cm, 6cm}, clip, width=\columnwidth, page=3]{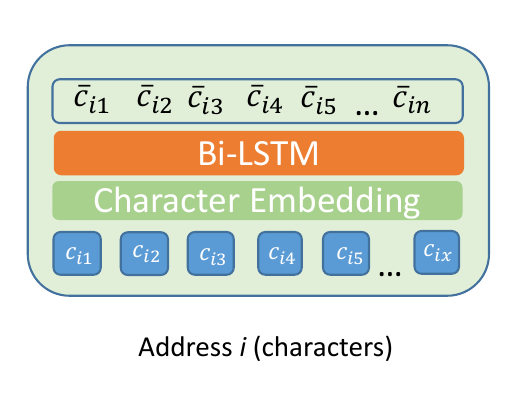}}
    
    \caption{\textbf{Character vector computation.} The computation of character vectors is very similar to that of word vectors, except pre-trained embeddings aren't used.}
    \label{fig:character_vectors}
\end{figure}

\begin{figure*}[hbt]
    \centering
    \makebox[\textwidth][c]{\includegraphics[trim={0cm, 1cm, 0cm, 1.5cm}, clip,width=1.2\textwidth, page=4]{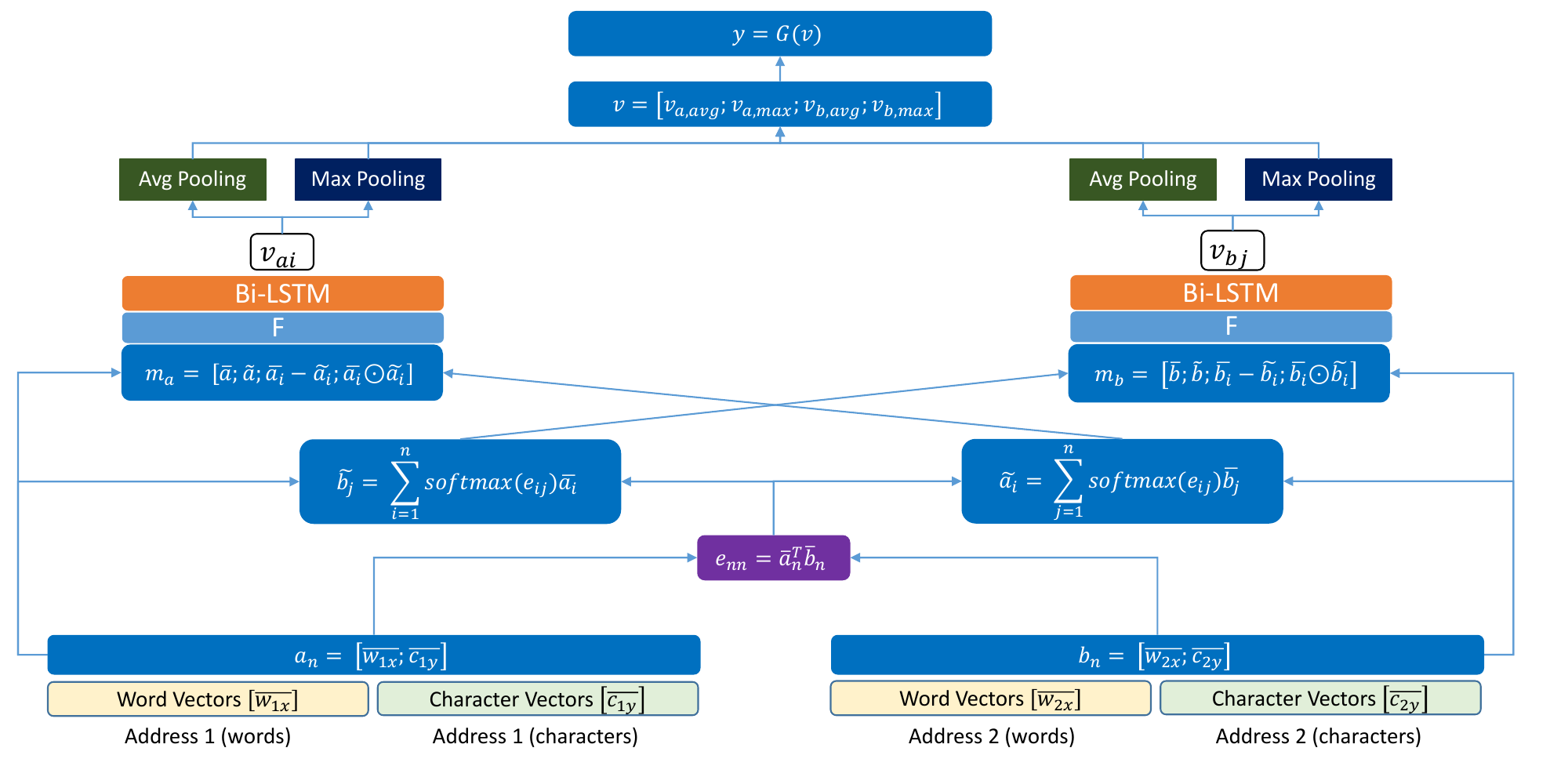}}
    
    \caption{\textbf{The modified ESIM architecture.} The original version of ESIM formulated by Chen \textit{et al.,} \cite{Chen_2017} did not contain the Character vectors and only worked with word vectors and embeddings. Dong \textit{et al} \cite{dong2018enhance} later studied the impact of adding character emebeddings, as shown in this figure in the context of next utterance selection in dialogues. We analyze the effectiveness of the ESIM model with and without the character embeddings for the task of address matching. The computations of word vectors and character vectors are shown in Figure~\ref{fig:word_vectors} and Figure~\ref{fig:character_vectors}.}
    \label{fig:architecure}
\end{figure*}

Thus, the soft alignment is obtained for each input string. This alignment vector is then concatenated with the respective input's combined embeddings, as well as the difference between them and their element-wise products. The rationale behind this step can be understood as trying to capture different transformations of the input words and characters. This concatenated vector for each input is later passed one more time through a BiLSTM layer, to capture sequential dependencies. Finally, the output of the BiLSTM layer is passed through two types of pooling operations. The model considers max and average pooling; and each of these operations generate a vector. The four output embeddings (Average and Max pool for each input) are then concatenated and passed through a fully connected layer to finally result in a binary softmax classification layer. 

Lin \textit{et al.}~\cite{lin20} show how the ESIM model~\cite{Chen_2017} can be adapted for the task of address matching. Chen \textit{et al.}~\cite{Chen_2017} have shown how this model can be used for sequential inference and, given the fact that addresses are essentially sequential bodies of text, their usage in this setting is pertinent. Lin \textit{et al.} ~\cite{lin20} were able to apply successfully this model in the context of Chinese address matching and showed that the model was effective.

We adapted this model for the usage of English language addresses as well, with some modifications. Instead of making the model directly consume the words in the addresses, we used pre-trained Glove embeddings~\cite{pennington2014glove} to supply the model with inputs as shown in Figure~\ref{fig:word_vectors}. The pre-trained Glove model we use generates 300 dimensional embeddings for every word in its vocabulary. These embeddings have been built based on co-occurence statistics of the tokens. The intuition here was that semantically sensible embeddings that capture the understanding of the similarity between typical words used in addresses (as the likes of those mentioned in Table~\ref{tab:examples}) could make it easier for the model to understand similar addresses. Further, while the initial formulation of ESIM only used word vectors, we also studied the effect of adding character embeddings to the original ESIM model (see Figure~\ref{fig:character_vectors}). There is one crucial difference to note here: while the character embeddings are initialized randomly using the Lecun Uniform initializer~\cite{lecuninit} and can be trained through the training process, the word embeddings are initialized with Glove and frozen through the training process. This is done to learn the character embeddings from scratch, and at the same time preserve the pretrained word embeddings.

Thus, the learning problem is formulated as follows. Let $x_1$ and $x_2$ be two addresses. We are tasked with learning the function.

\begin{equation}
    F(<x_1, x_2>) = y
\end{equation}

where $y=1$ if $x_1$ and $x_2$ are a match and $y = 0$ otherwise.

\section{Experiments and Results}
\label{sec:experiments}

The above proposed models were evaluated along three measures: Precision, Recall and Accuracy. Most of these models do not require to be trained and can be directly employed on a set of addresses. The deep learning based ESIM model however requires to be trained and validated. Therefore, the previously mentioned training and development data was used to train the ESIM model (with and without the character embeddings) and the test data was used in the evaluation of all approaches.

It might be worthwhile to discuss the training process of the ESIM + Char Embedding model. The training parameters for the model are provided in Table~\ref{tab:final_model}. Originally, the authors had used the Adam optimizer for the ESIM model~\cite{Chen_2017}, and therefore we decided to use the same one.  The training was conducted on an AWS g4dn.12xlarge machine, and early stopping was employed to make sure the model does not train excessively in scenarios where they do not tend to converge. We studied the training process for various learning rates, namely $10^{-2}$, $10^{-3}$, and $10^{-4}$. For the sake of brevity, we are not discussing the hyper-paramter tuning performed on the original ESIM model, but it should be noted that the same process was applied to that model as well.

\begin{table}[hbt]
\centering
\begin{tabular}[t]{lr}
\toprule
Parameter&Value\\
\midrule
Optimizer&Adam\\
Learning Rate&1e-4\\
Loss&Binary Crossentropy\\
Patience (ES)&4\\
Max. Epochs&50\\
Batch Size&8\\
\bottomrule\\
\end{tabular}
\caption{Training Parameters for the ESIM + Character Embedding model}
\label{tab:final_model}
\end{table}%

Table~\ref{tbl:apr_curves} shows the curves for important measures like Accuracy, Precision, Recall and Loss while the model is being trained across these different learning rates. These measures are shown for both training and validation sets. For a learning rate of $10^{-2}$, the loss curve suggests that the model has not even converged, and this is largely attributed to the fact that a consistent decrease in training loss is not observed. For a learning rate of $10^{-3}$ there is sufficient evidence to suggest model convergence, from the loss curves. However, further analysis of the precision, recall and accuracy curves does not suggest a consistent value and therefore, this learning rate still does not seem to be suitable. This is especially the case with the accuracy and recall values as they fluctuate greatly for both test and validation data and do not converge. A learning rate of $10^{-4}$, however, provides the best performance for the model in terms of the loss curves and evaluation metrics. The model seems to have converged from the loss curves, and the accuracy, precision and recall measures are consistent. Hence, we train the model with a learning rate of $10^{-4}$. From table~\ref{tbl:apr_curves}, the point of overfitting for this learning rate can be observed clearly, demarcated by the red line.

\begin{table*}[hbt]
    \centering
        \begin{tabular}{|@{}c@{}|@{}c@{}|@{}c@{}|@{}c@{}|@{}c@{}|@{}c@{}|}
           \toprule
            Metric/$lr$ & $10^{-2}$ & $10^{-3}$ & $10^{-4}$ & $10^{-5}$ \\
            \midrule
            \centered{Loss} & \includegraphics[scale=0.255]{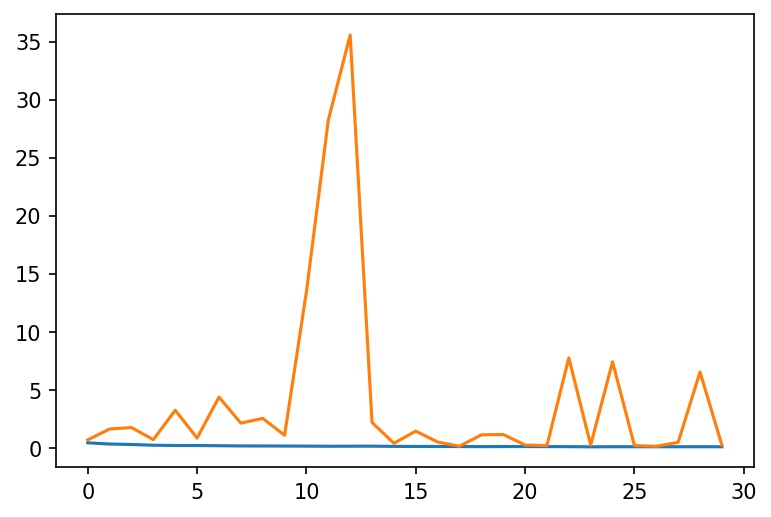} &   \includegraphics[scale=0.255]{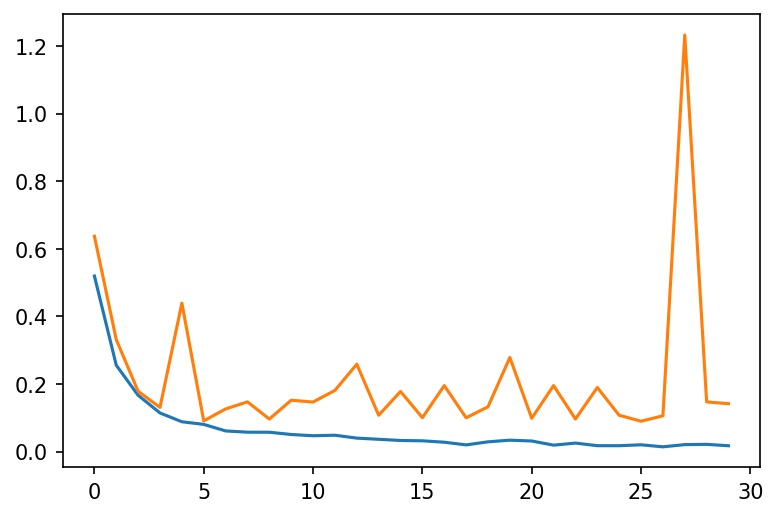} &   \includegraphics[scale=0.122]{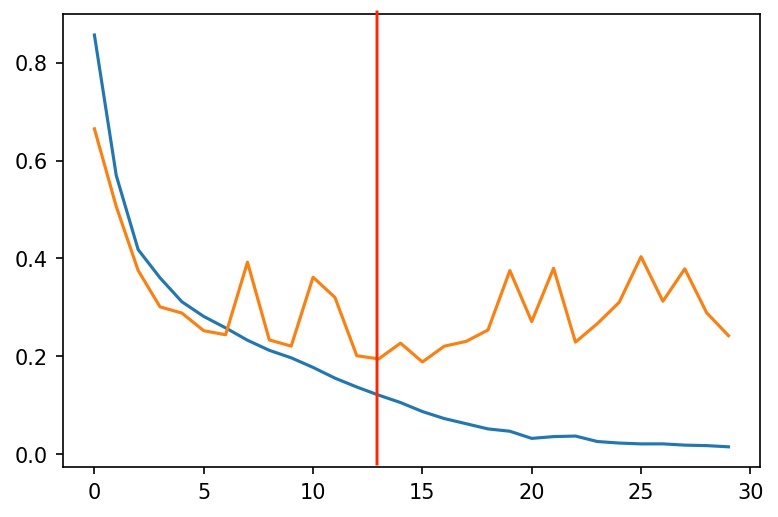} &   \includegraphics[scale=0.255]{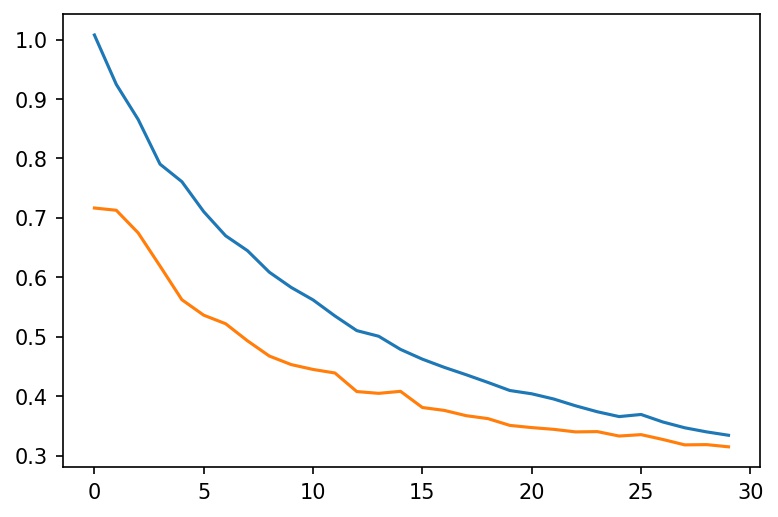} \\
            \centered{Accuracy} & \includegraphics[scale=0.255]{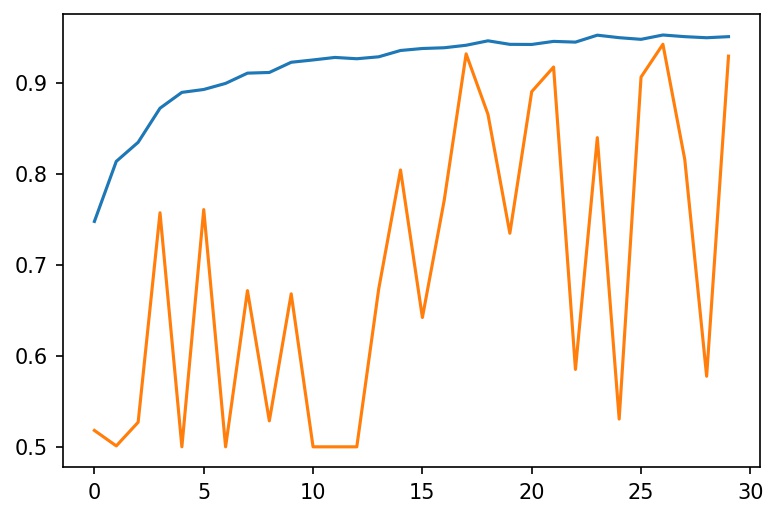} &   \includegraphics[scale=0.255]{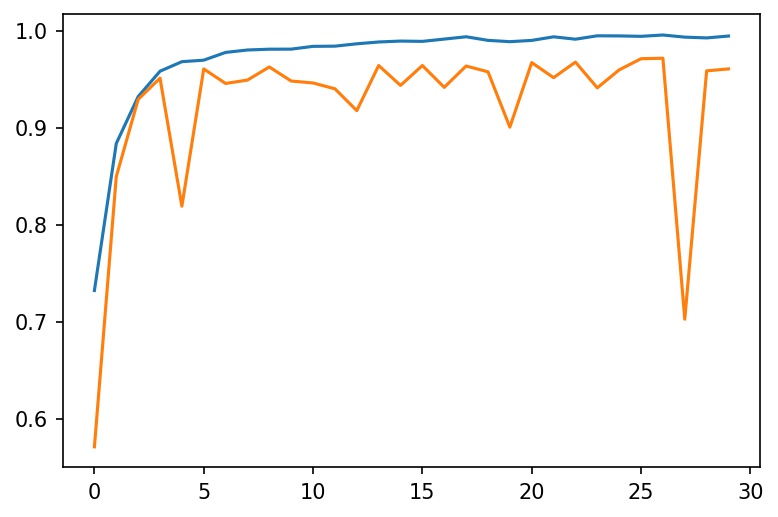} &   \includegraphics[scale=0.255]{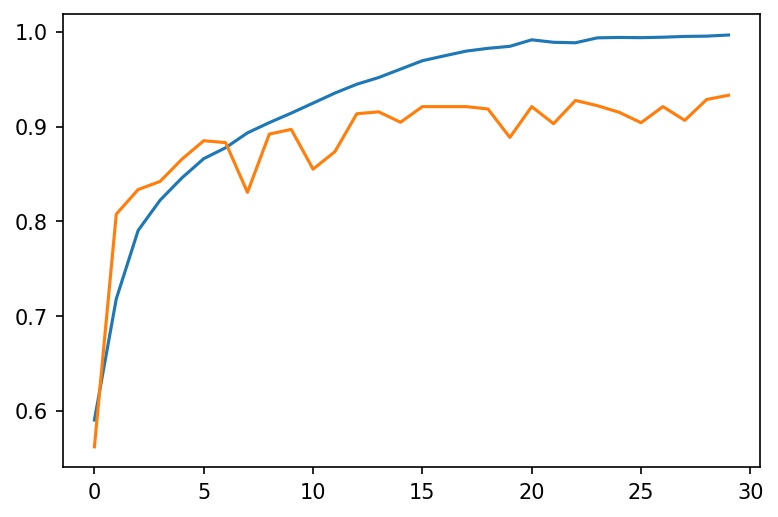} &   \includegraphics[scale=0.255]{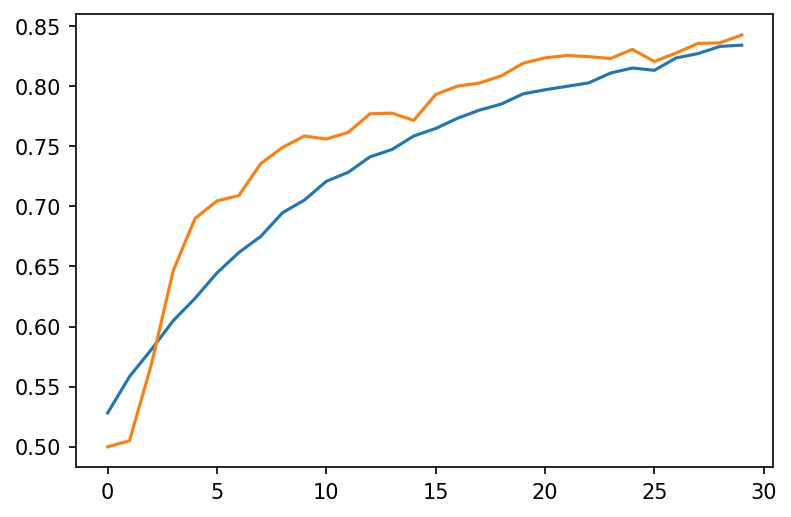} \\
            \centered{Precision} & \includegraphics[scale=0.255]{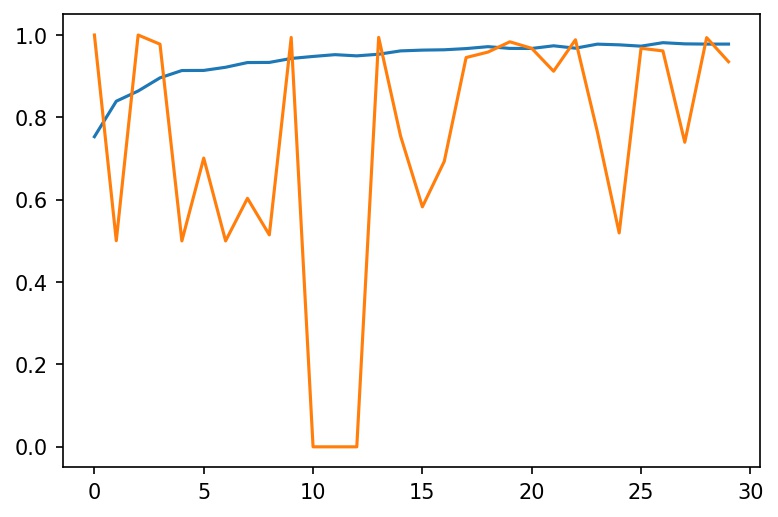} &   \includegraphics[scale=0.255]{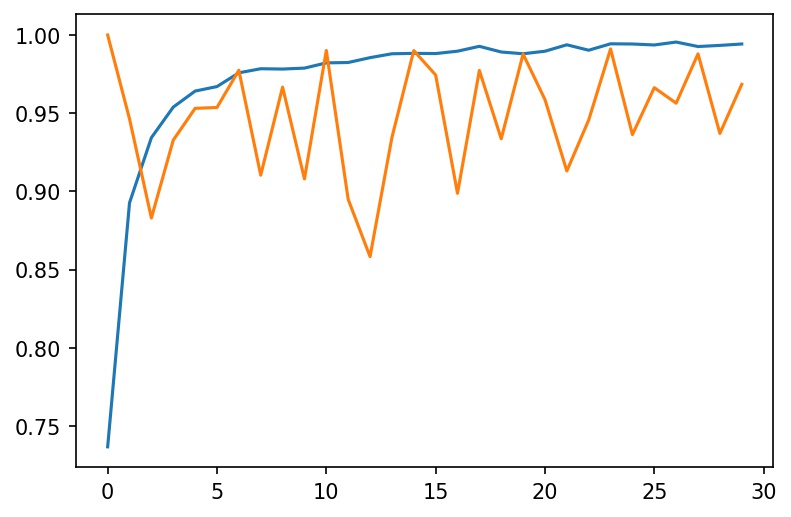} &   \includegraphics[scale=0.255]{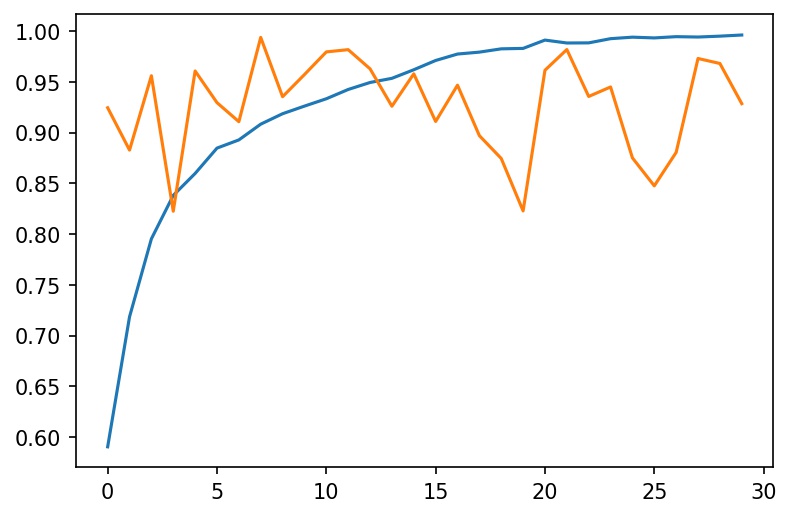} &   \includegraphics[scale=0.255]{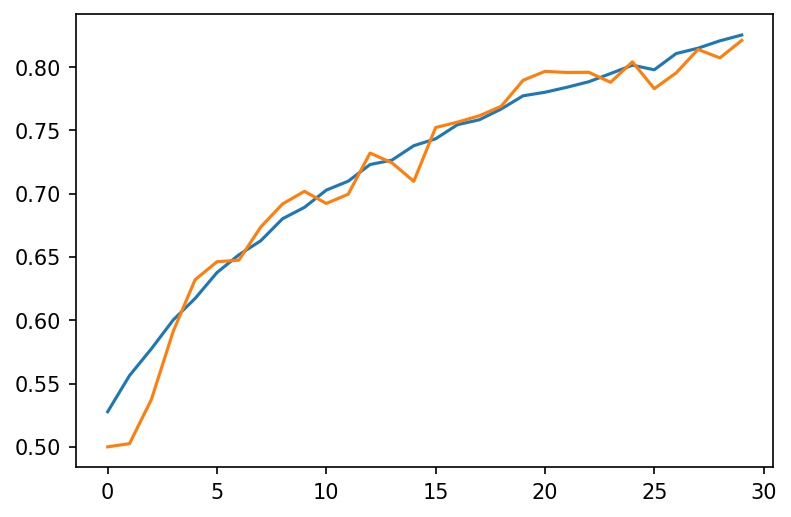} \\
            \centered{Recall} & \includegraphics[scale=0.255]{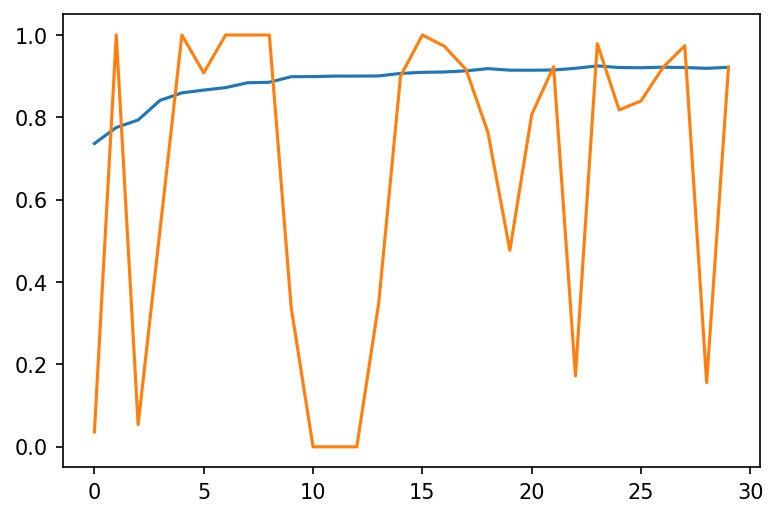} &   \includegraphics[scale=0.255]{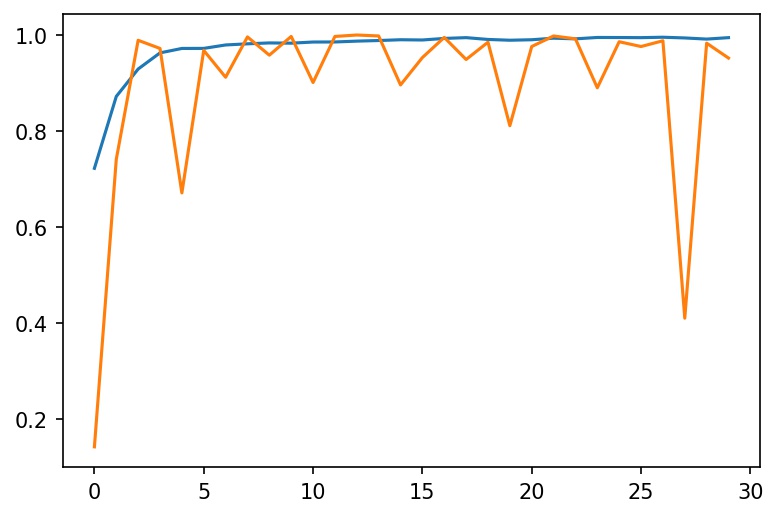} &   \includegraphics[scale=0.255]{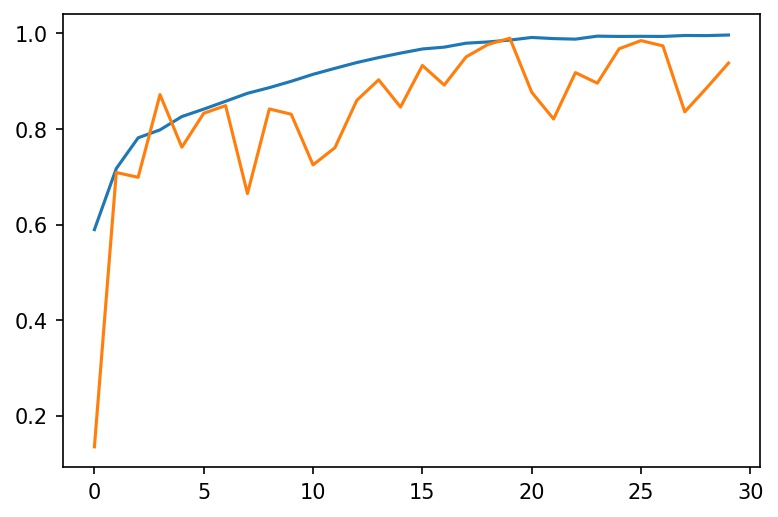} &   \includegraphics[scale=0.255]{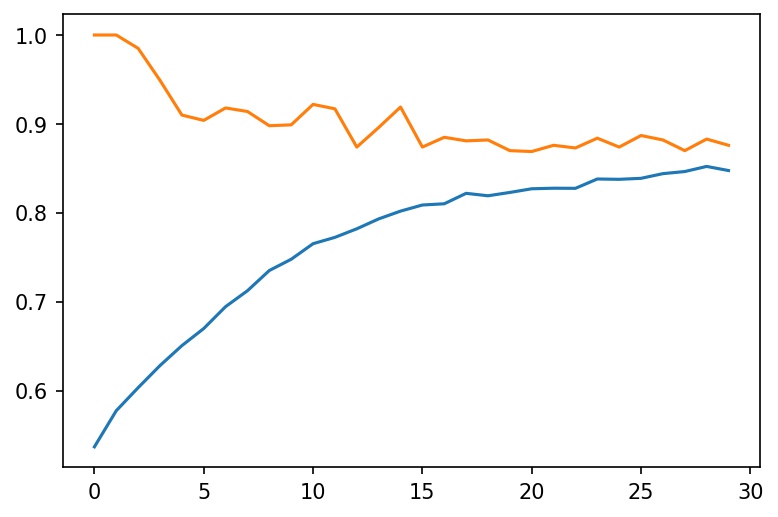} \\
            \bottomrule
        \end{tabular}
    \caption{The performance of the ESIM + Char embedding model across three different learning rates when evaluated against  Loss, Accuracy, Precision and Recall. The orange line represents the performance of the model on the training data and the blue line represents the performance of the model on the validation data. In the case of the loss curve for learning rate of $10^{-4}$, the point of overfitting can be observed clearly (red line).}
    \label{tbl:apr_curves}
\end{table*}

Table~\ref{tab:performance} shows the results that were obtained for all the models. The models can be grouped into three broad categories - high precision, high recall and overall performance.

\begin{table*}[hbt]
    \centering
    \begin{tabular}{lrrrr}
%    \begin{tabularx}{\textwidth}{WZZZZ}
        \toprule
        \textbf{Model} & \textbf{Precision} & \textbf{Recall} & \textbf{Accuracy} & \textbf{Time (s)}\\ 
        \midrule 
        plain & {\bf0.98} & 0.24 & 0.62&0.46\\ %\midrule
        segment & 0.94 & 0.69 & 0.82 &125.64\\ %\midrule
        normalized-plain & {\bf0.98} & 0.24 & 0.62&0.47\\ %\midrule
        tokens-jacquard & 0.96 & 0.16 & 0.58&{\bf0.10}\\ %\midrule
        segment-tokens-jacquard & 0.96 & 0.57 & 0.77&104.17\\ %\midrule
        n-grams-jacquard & 0.53 & {\bf1.00} & 0.55&0.30\\ %\midrule
        segment-n-grams-jacquard & 0.88 & 0.87 & 0.88&104.35\\ %\midrule
        levenshtein & 0.70 & 0.22 & 0.56 &3.31\\ %\midrule
        segment-levenshtein & 0.96 & 0.78 & 0.87 &104.51\\ %\midrule
        jaro-winkler & 0.66 & 0.58 & 0.64 & 0.30\\
        segment-jaro-winkler & 0.78 & 0.94 & 0.84 & 107.11\\
        tfidf & 0.60 & 0.39 & 0.57 &40.41\\ %\midrule
        segment-tf-idf & 0.59 & 0.96 & 0.65 &136.12\\ \midrule
        ESIM - Train & 0.99\color{black}(0.01)\color{black} & 0.99\color{black}(0.00)\color{black} & 0.99\color{black}(0.00)\color{black} & 533.11\color{black}(39.30)\color{black}\\
        ESIM - Test & 0.93\color{black}(0.02)\color{black} & 0.95\color{black}(0.02)\color{black} & 0.94\color{black}(0.01)\color{black} & 0.74\color{black}(0.04)\color{black}\\
        ESIM + Char - Train & 0.99\color{black}(0.00)\color{black} & 0.99\color{black}(0.00)\color{black} & 0.99\color{black}(0.00)\color{black} & 827.85\color{black}(65.60)\color{black}\\
        ESIM + Char - Test & 0.95\color{black}(0.01)\color{black} & 0.94\color{black}(0.00)\color{black} & {\bf 0.95\color{black}(0.00)\color{black}}&1.05\color{black}(0.02)\color{black}\\
        \bottomrule
    \end{tabular}
    \caption{Test 4. Performance of different models. Since the ESIM based models depend upon the random initialization of the layers we, re these evaluations for 5 different random seeds. The mean and standard deviation for each metric has been reported.}
    \label{tab:performance}
\end{table*}

It can be seen that most approaches that involve looking at a distance metric in a non segmented fashion like plain, normalized-plain, tokens-jacquard and levenshtein have a relatively higher precision and suffer in terms of recall. This can be explained by the strict definition of matches that such approaches impose. These models might not be able to capture the variety of noise that has been injected into matches and since they are low on expressivity. The tfidf approach behaves in the aforementioned way too. This can be attributed to the strictness of the match imposed by the character level 3-grams.

In terms of the high recall models, n-grams-jacquard has a perfect recall, but a poor precision. This signifies that this approach cannot tackle nuances in mismatches and tends to classify more records as matches than not.

Segmentation seems to have a positive effect overall, as it helps some distance based approaches achieve the best of both worlds. segment-tokens-jacquard improves the recall when compared to tokens-jacquard; segment-n-grams-jacquard balances out the precision and recall when compared to n-grams-jacquard; segmented-levenshtein improves the precision and recall from levenshtein and finally segment-tf-idf improves the recall of tf-idf. One can clearly see how segmentation is improving the results, and it can be related to this approach's capability of treating matches and mismatches at a field level independently.

The ESIM + Character embeddings model however seems to have the best overall performance. It has the highest value for accuracy, and this can be alluded to the capacity of the model to understand various notions of similarity. While the accuracy numbers are great for this model, its only shortcoming is that it requires significantly higher time to train. Another caveat is that this model is strictly based off the training data. Hence, if we are to match addresses of a different granularity, this model will be brittle. It must also be noted that the original ESIM model marginally falls short of the ESIM + character embeddings model (in terms of accuracy and precision). This can be attributed to the character embeddings providing the necessary infrastructure needed to deal with character noise, that the word level embeddings alone will not be able to handle.

Since the ESIM + Char model marginally outperforms the ESIM model, it would be safe to say that this model has the best overall performance out of all the models we have considered. However, we need to understand the generalization capability of this model to real world addresses (as the dataset it was tested on was synthetically generated). In order to understand this, we built a manually annotated dataset of addresses from the Companies House dataset\footnote{http://download.companieshouse.gov.uk/en\_output.html}. 100 pairs of addresses were built from this dataset and labels for matching addresses at the building level granularity were assigned. Out of the 100 pairs, there were three that were matching and the rest 97 were mismatches. On this dataset, the ESIM + Char embedding model was able to achieve an accuracy of 71.8\% (20.64), precision of 15.30\% (11.64) and recall of 80\% (26.66). This performance suggests that the deep learning model is able to perform well as a high recall classifier in domains that it was not trained on. However, reduction of the standard deviation of the metrics (model stability) in out of sample data is potential future work.

While the deep learning based models have demonstrated great competitiveness in terms of Accuracy, Precision, and Recall, one shortcoming is the amount of time taken for training. However, it is also important to note that this is only the case when we train the model. During test time however, they take less than 2 seconds to produce the results - which is comparable to most approaches. These values are expected to be greater when they are trained on systems that do no have a GPU. 

\section{Conclusions}
\label{sec:conclusions}

The main contributions of our work are as follows:
\begin{itemize}
    \item A framework to generate a robust dataset of matching and mismatching English language addresses. These were based on a set of operations that we observed from real life addresses.
    \item An adaption of a deep learning architecture from the domain of response generation in dialogues\reviewerone{,} to the address matching problem.
    \item A comparative study of various approaches to perform the task of address matching. 
\end{itemize}

The data generation process was able to capture various nuances and subtleties present in working with addresses. These transformations sufficiently demonstrate how different and unique addresses are, within the world of textual data. 

Empirically, we were able to arrive at the conclusion that on the generated dataset, the ESIM + Character Embeddings model was able to achieve the overall best performance in terms of Precision, Recall and Accuracy. This can be attributed to the model's capacity of capturing the variations that occur in the matches and mismatches. Though other approaches do not match the overall performance of this model, they compare reasonably well (especially the segment-n-grams-jacquard and segment-levenshtein methods) and are comparably faster. For the past few years there have been stunning advancements in the field of Natural Language Processing. Specifically, with the advent of BERT~\cite{devlin2018bert} the landscape of deep learning in NLP has changed dramatically. Since attention is an integral part of BERT, we believe that the architecture is well suited for the task of address matching. Therefore, one potential future direction will be to assess the effectiveness of models like BERT with an increase/decrease in the level of granularity. Further, instead of Glove being used as the word embeddings, BERT can also be plugged in and adapted to this framework. Making these embeddings trainable would also yield address relevant BERT embeddings/Glove embeddings.

Potential areas of improvement include, but are not limited to, improving realism in the data generation process, and altering the setting to study more fine/coarse grained address matching, etc. While the ESIM model has shown its capabilities of being a strong foundation for the task of address matching, one could also probe into performing a more detailed hyperparameter tuning of the ESIM model or alter the architecture to improve the evaluation metrics. As suggested in the previous section, improvement of model stability in a different domain of addresses to where the model was trained is also a great avenue for future work to explore.

\section{Acknowledgements}
We would like to acknowledge and thank the inputs of Mark Hunt and Omar Cresdee which helped us enrich our work.
\newline
\newline
This paper was prepared for informational purposes by the Artificial Intelligence Research group of JPMorgan Chase \& Co and its affiliates (“J.P. Morgan”), and is not a product of the Research Department of J.P. Morgan.J.P. Morgan makes no representation and warranty whatsoever and disclaims all liability, for the completeness, accuracy or reliability of the information contained herein.This document is not intended as investment research or investment advice, or a recommendation, offer or solicitation for the purchase or sale of any security, financial instrument, financial product or service, or to be used in any way for evaluating the merits of participating in any transaction, and shall not constitute a solicitation under any jurisdiction or to any person, if such solicitation under such jurisdiction or to such person would be unlawful.
\newline
\newline
© 2022 JPMorgan Chase \& Co. All rights reserved

%\clearpage

\printbibliography
\end{document}